\documentclass[aps,pra,preprint,graphicx,superscriptaddress,longbibliography,onecolumn]{revtex4-1}
\usepackage{graphicx,amsfonts}
\usepackage{amsmath}

\begin{document}


\title{Intense high harmonic vector beams from relativistic plasma mirrors} 



\author{Zi-Yu Chen}
\email[]{ziyuch@scu.edu.cn}
\affiliation{Key Laboratory of High Energy Density Physics and Technology (MoE), College of Physics, Sichuan University, Chengdu 610064, China}
\affiliation{National Key Laboratory of Shock Wave and Detonation Physics, Mianyang 621999, China}
\author{Ronghao Hu}
\email[]{ronghaohu@scu.edu.cn}
\affiliation{Key Laboratory of High Energy Density Physics and Technology (MoE), College of Physics, Sichuan University, Chengdu 610064, China}

%


\date{\today}

\begin{abstract}
Vector beam, with a spatial nonuniform polarization distribution, is important for many applications due to its unique field characteristics and novel effects when interacting with matter. Here through three-dimensional particle-in-cell simulations, we demonstrate that intense vector beams in the extreme-ultraviolet to x-ray spectral region can be generated by means of high harmonic generation (HHG) in the relativistic regime. The vector features of the fundamental laser beam can be transferred to the higher frequency emission coherently during the extreme nonlinear HHG dynamics from relativistic plasma mirrors. The vector harmonic beams can be synthesized into attosecond vector beams. It is also possible to generate vector harmonic beam carrying orbital angular momentum. Such bright vortices and vector light sources present new opportunities in various applications such as imaging with high spatial and temporal resolution, ultrafast magnetic spectroscopy, and particle manipulation.
\end{abstract}


\maketitle 

\section{Introduction}
Polarization as a fundamental property of light plays a critical role in determining the interaction between light and matter. The commonly known states of polarization, such as linear polarization, circular polarization, and elliptical polarization, refer to spatially uniform light beams, i.e., the states of polarization are the same at different points of the transverse polarization plane. In comparison, vector beams are light beams with spatially non-uniform state of polarization\cite{zhan_cylindrical_2009}. Radially and azimuthally polarized light beams are two typical examples. At each point in the polarization plane, the electric field is linearly polarized locally; however, the field vectors point along different directions. 

Generation of vector beams has attracted growing interest recently, as they exhibit unique characteristics and novel effects when interacting with matter. For example, they can be tightly focused beyond the diffraction limit and exhibit strong non-vanishing longitudinal electric or magnetic fields\cite{quabis_focusing_2000,dorn_sharper_2003,hnatovsky_revealing_2011}. These interesting properties can be explored for a number of applications involving polarization-sensitive imaging\cite{chen_imaging_2013}, optical trapping\cite{zhan_trapping_2004,kozawa_optical_2010}, high-capacity communication\cite{milione_using_2015,chen_vector_2020}, laser micromachining\cite{meier_material_2007,drevinskas_laser_2016}, and relativistic particle acceleration\cite{salamin_direct_2008,zaim_relativistic_2017,wen_electron_2019}. 

Though various methods, both active and passive, e.g., tunable q-plates\cite{rumala_tunable_2013,dambrosio_arbitrary_2015,larocque_arbitrary_2016}, can be employed to generate vector beams in the infrared and visible spectral region, it is difficult to extend these linear optics based techniques to generate vector beam in the extreme-ultraviolet (XUV) and x-ray regime that would further extend its power and capability. Coherent short-wavelength light sources would allow a better spatial resolution in imaging, which is often limited by the wavelength of the light. Combining the unique properties of vector beams, e.g., focused azimuthally polarized beams exhibiting strong longitudinal magnetic field, XUV/x-ray vector beams can be a powerful tool in applications such as imaging and control of magnetic properties in the nanoscale. Recently, spatial-polarization independent parametric up-conversion of vector beams has been realized\cite{liu_nonlinear_2018,wu_spatial-polarization-independent_2020}; however, only second harmonic in the perturbative nonlinear optics regime is obtained. In the nonperturbative extreme nonlinear optics regime, Hern\'andenz-Garc\'ia et al. demonstrated the generation of radially and azimuthally polarized XUV vector beams via high harmonic generation (HHG) from infrared driving vector beams interaction with gases target\cite{hernandez-garcia_extreme_2017}. This approach has the advantage of generating ultrafast vector beams in the attosecond regime and easily controlling the beam properties by modifying the driving beam instead of using inefficient XUV optical components. Yet, intrinsic to the generation mechanism of gas HHG, i.e., the so-called three-step model of ionization-acceleration-recombination\cite{corkum_plasma_1993}, the driving laser intensity is limited (typically $ 10^{14} $ W/cm$ ^2 $) and thus the yield of vector high harmonics is relatively low.

To get brighter ultrafast XUV/x-ray sources, HHG from relativistically intense (intensities $ \geq 10^{18} $ W/cm$ ^2 $ for wavelength of about 1 $ \mu $m) laser-driven oscillating overdense plasma surfaces (termed as plasma mirrors) has been demonstrated as a promising route. The underlying mechanism is generally described as relativistically oscillating mirrors (ROM) \cite{bulanov_interaction_1994,lichters_shortpulse_1996,baeva_theory_2006,dromey_high_2006,pukhov_x-rays_2006,dromey_bright_2007,thaury_plasma_2007}. While HHG from ROM with controllable polarization\cite{chen_bright_2016,ma_intense_2016,chen_isolated_2018,blanco_controlling_2018,chen_isolated_atto_2018,chen_spectral_2018} (i.e., beyond linear polarization) and vortex wavefront\cite{zhang_generation_2015,denoeud_interaction_2017,leblanc_plasma_2017} (i.e., with orbital angular momentum, OAM) has been investigated intensively, HHG with vector beam characteristics has not been studied yet. Very recently, Za\"im ea al. reported the first HHG in the relativistic regime using radially and azimuthally polarized laser pulses\cite{zaim_interaction_2020}. However, only the angularly resolved harmonic spectra with respect to harmonic intensity are considered. No measurement with respect to the spatial polarization distribution of the high harmonics has been done. Whether high harmonic vector beams can be generated remains elusive.

In this work, through three-dimensional (3D) particle-in-cell (PIC) simulations, we numerically demonstrate the generation of intense vector beams in the XUV regime via HHG from relativistic plasma mirrors. The electric field-vector distribution shows that the driving laser pulse imprints its vector beam pattern into the higher frequency radiation during the extreme nonlinear process of HHG based on ROM. The vector harmonic beams can be generated with a high efficiency and synthesized into attosecond vector beams. Besides, we show that high harmonic vector beams with nonzero OAM can be generated under oblique incidence.

\section{3D Simulation setup}

The 3D PIC simulations are performed using the code VLPL (Virtual Laser Plasma Lab)\cite{pukhov_three-dimensional_1999}. The simulation box size is $ 16\lambda_L \times 12\lambda_L \times 12\lambda_L $ for normal incidence and $ 20\lambda_L \times 20\lambda_L \times 20\lambda_L $ for oblique incidence, where  $ \lambda_L = 800 $ nm is the wavelength of the driving laser beam. The grid step size is $ \lambda_L/40 $, $ \lambda_L/25 $, and $ \lambda_L/25 $ in the $ x$-, $y$-, $z $-direction, respectively. Each cell is filled with 8 macroparticles. The temporal laser profile has a Gaussian envelope $ L_T=a_0 \exp(-t^2/2\tau_0^2) $, where $ a_0 = eA/m_e c^2 = 5 $ is the normalized laser amplitude (corresponding to a laser intensity of $ I_0 = 5.4\times 10^{19} $ W/cm$ ^2 $), $ \tau_0 = 2T_0 $, $ T_0 $ is the laser period, $ A $ is the laser vector potential, $ e $ is the elementary charge, $ m_e $ is the electron mass, and $ c $ is the light speed in vacuum. We set the incident laser pulses propagating along the $x$-axis and polarized in the $y-z$ plane.

For the transverse modes, three kinds of cylindrically symmetric vector beams are considered for demonstration in this study, i.e., (1) a radially polarized pulse $ \textbf{E}_r $; (2) an azimuthally polarized pulse $ \textbf{E}_{\phi} $; (3) a generalized vector beam $ \textbf{E}_g $. In our simulations, we make use of superposition of orthogonally polarized Hermite-Gauss modes to generate these vector beams:
\begin{equation}\label{rp}
	\textbf{E}_r = \mathrm{HG}_{l=1,m=0} \textbf{e}_y + \mathrm{HG}_{l=0,m=1} \textbf{e}_z,
\end{equation}
\begin{equation}\label{ap}
	\textbf{E}_{\phi} = \mathrm{HG}_{l=0,m=1} \textbf{e}_y - \mathrm{HG}_{l=1,m=0} \textbf{e}_z,
\end{equation}
\begin{equation}\label{gv}
	\textbf{E}_g = \textbf{E}_r + \textbf{E}_{\phi},
\end{equation}
where $ \textbf{e}_y $ and $ \textbf{e}_z $ is the unit vector along the $ y $- and $ z $-axis, respectively. Hermite-Gaussian modes are the paraxial beam-like solutions to the scalar Helmholtz equation in Cartesian coordinates. The expression for the Hermite-Gaussian modes is given by\cite{siegman1986}
\begin{equation}\label{hg}
	\mathrm{HG}_{l,m} = u_0 u_l(x,y) u_m(x,z) \exp(-i\textbf{k} \cdot \textbf{x} + i\varphi_0),
\end{equation}
with 
\begin{equation}\label{key}
\begin{split}
	u_J(x,s) = & \bigg(\frac{\sqrt{2/\pi}}{2^J\, J!\, w_0}\bigg)^{1/2} \bigg( \frac{q_0}{q(x)} \bigg)^{1/2} \bigg(- \frac{q^{\ast}(x)}{q(x)} \bigg)^{J/2} \\
&	\times H_J \bigg( \frac{\sqrt{2}s}{w(x)} \bigg) \, \exp \bigg( -i \frac{ks^2}{2q(x)} \bigg),
\end{split}
\end{equation}
where $ q(x) = x + i x_{R} $ is the complex beam parameter and $ q^{\ast} $ is its complex conjugate, $ x_{R}=\pi w_0^2 / \lambda_L $ is the Rayleigh length for the beam of waist $ w_0=2\lambda_L $, $ H_J(\sqrt{2}s/w(x)) $ is the Hermite polynomial of order  $ J $, $ w(x) = w_0\sqrt{1+(x/x_{R})^2}$ is the beam radius at position $ x $, $ \textbf{k} $ is the wave vector, $ \varphi_0 $ is initial phase, $ u_0 $ is a normalizing factor, $ J=l,m $, and $ s=y,z $. For $ l=m=0 $, $ \mathrm{HG}_{0,0} $ reduces to the fundamental Gaussian beam. 

The transverse electric field of a radially polarized laser pulse can also be expressed as\cite{esarey_laser_1995}
\begin{equation}\label{key}
	\textbf{E}_r = E_0 \frac{\textbf{r} w_0}{w(x)^2} \exp(-r^2/w^2) \sin \psi,
\end{equation}
where 
\begin{equation}\label{key}
	\psi = kx + x r^2 / (x_R w^2) -2 \arctan(x/x_R)+\varphi_0.
\end{equation}
From this expression, it can be seen that the laser amplitude at focus reaches its peak of $a=a_0/\sqrt{2e}\approx 0.42a_0$ when $r=w_0/\sqrt{2}$, smaller than the peak field amplitude $ a_0 $ of the Gaussian beams. 

Experimentally, these relativistically intense vector laser beams can be produced by converting a linearly polarized laser pulse into the desired polarization structures using a phase mask consisting of space-variant waveplates with different optical axes before focusing to high intensity, as demonstrated by Za\"{i}m et al\cite{zaim_interaction_2020}.

\section{Results and discussion}
\subsection{Vector laser beam}
\begin{figure*}[htbp]
	\centering
	\includegraphics[width=0.8\textwidth
	]{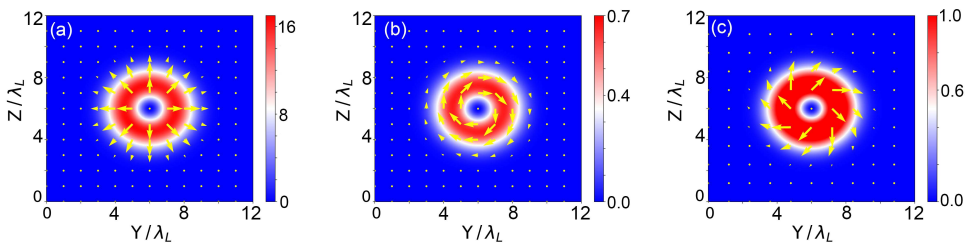}
	\caption{\label{laser} Electric filed distribution of the incident vector laser beams obtained from 3D PIC simulations. (a) The radially polarized laser $ \textbf{E}_{r} $. (b) The azimuthally polarized laser $ \textbf{E}_{\phi} $. (c) The generalized vector laser beam $\textbf{E}_g = \textbf{E}_r + \textbf{E}_{\phi}$. The color scale shows the intensity pattern in the transverse plane, while the yellow arrows represent the instantaneous local electric field vectors.}
\end{figure*}

The spatial distribution of instantaneous electric field for the radially polarized, azimuthally polarized, and generalized vector laser beams obtained from the PIC simulations are graphically illustrated in Fig. \ref{laser}(a)-(c), respectively. The color scale shows the intensity pattern, while the yellow arrows represent the characteristics of the spatial polarization distribution. All the three vector beams feature a donut-shaped intensity profile with a polarization singularity in its center, as a result of transverse field continuity. 

\subsection{Vector harmonic beam generation}
To verify the generation of vector beams in the XUV/X-ray spectral region through HHG process from relativistic plasma surfaces, we first consider the case of vectorial infrared laser beams normally incident onto plasma surfaces. The plasma slab has a thickness of 0.5$ \lambda_L $ ($x=[11.5,12.0]\lambda_L$) and density of $ n_0 = 100 n_c $, where $ n_c=m_e \omega_0/4\pi e^2 $ is the plasma critical density with respect to the fundamental laser frequency $ \omega_0 $. In the front of the plasma slab, there exists a preplasma region with exponential density profile $ n_e(x) = n_0 \exp((x-x_s)/L_s) $, where $ L_s=0.2\lambda_L $ is the density scale length and $ x_s $ denotes the position of the plasma-slab front surface.

\begin{figure*}[htbp]
	\centering
	\includegraphics[width=0.8\textwidth
	]{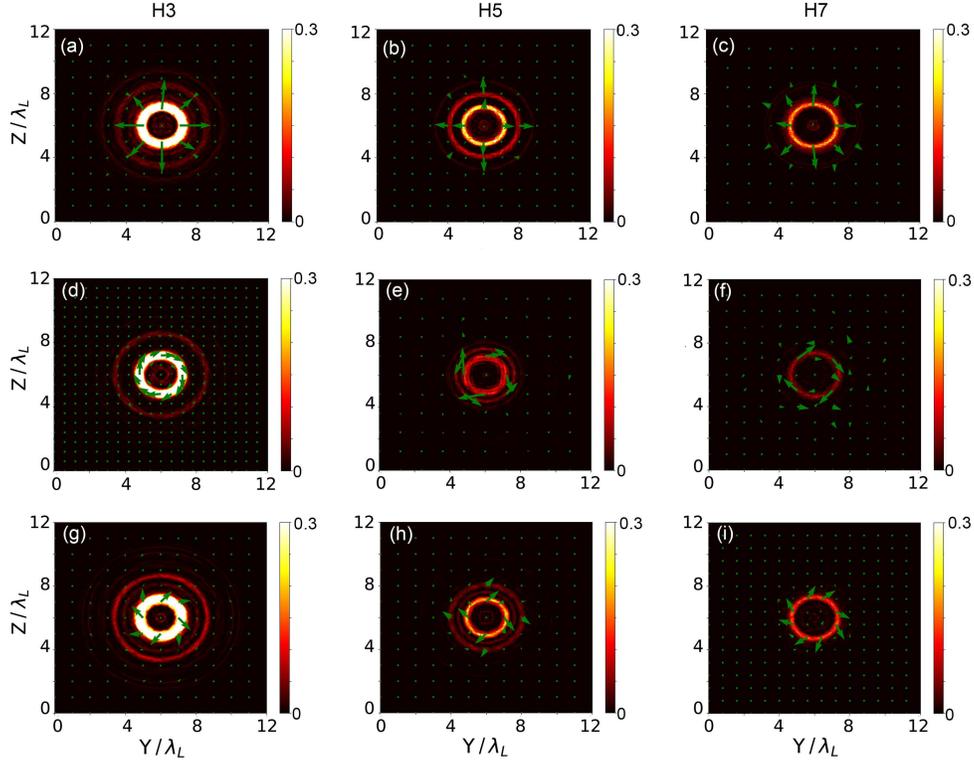}
	\caption{\label{hhg_nor} Electric filed distribution of the generated harmonics driven by the (a-c) radially polarized, (d-f) azimuthally polarized, and (g-i) generalized vector laser beams. The lasers are normally incident onto the plasma surfaces. The observation planes are at $ x=6.2\lambda_L $. The first to third columns correspond to the third (H3), fifth (H5) and seventh (H7) harmonics, respectively. The color scale show the harmonic intensity patterns, with the green arrows denoting the instantaneous local electric field vectors.}
\end{figure*}

After the driving laser beam is completely reflected from the plasmas, harmonic fields can be obtained by analyzing the reflected electric fields with Fourier transform. Each harmonic with order $ n $ is selected by spectral filtering in the frequency range $ [ n-0.5,n+0.5 ] \omega_0 $. Figure \ref{hhg_nor} presents the results of harmonic intensity pattern and spatial polarization distribution in the transverse $ y-z $ plane at $ x=6.2\lambda_L $ for the 3rd (H3), 5th (H5), and 7th (H7) harmonics generated by the three kinds of vector laser beams. A striking feature with respect to the spatial polarization distribution can be clearly appreciated from the green arrows,  representing electric field vectors in Fig. \ref{hhg_nor}, that the vector beam patterns are coherently imprinted in each harmonic order, showing the polarization characteristics of the fundamental beam can be preserved during the extreme nonlinear HHG dynamics via ROM.  

For harmonic intensity, radially polarized lasers generate the most intense high harmonics (see Figs. \ref{hhg_nor}(a)-(c)), while HHG driven by azimuthally polarized lasers is the weakest among the three kinds of lasers. This is because radially polarized beams exhibit strong longitudinal electric fields when tightly focused, which is favorable for HHG based on the ROM mechanism. In contrast, azimuthally polarized beams only present strong longitudinal magnetic fields. The strength of longitudinal electric field and thus the harmonic intensity for the generalized vector beam fall in between the cases of radially and azimuthally polarized beams.

\begin{figure}[htbp]
	\centering
	\includegraphics[width=0.7\textwidth
	]{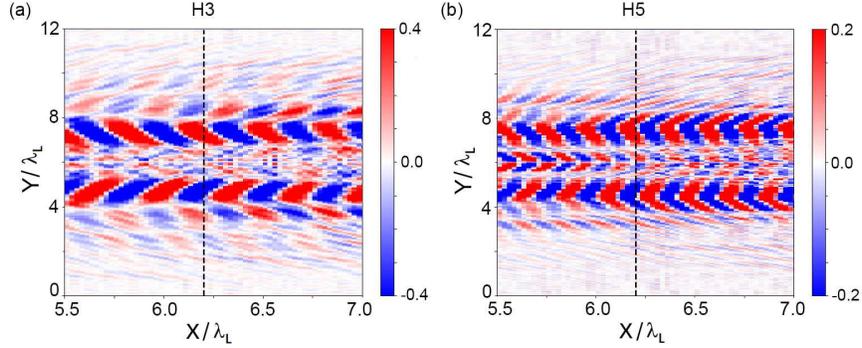}
	\caption{\label{high_order_mode} Higher-order transverse modes of the harmonics. Electric field distribution in the $ x-y $ plane for the (a) third and (b) fifth harmonics driven by radially polarized laser pulses. The black dashed lines at $ x=6.2 \lambda_L $ denote the observation positions in Fig. \ref{hhg_nor}.}
\end{figure}

Interestingly, the harmonics show spatial intensity modulation with a spherical mainlobe and a few sidelobes.  This intensity feature implies higher transverse modes with larger $ (l,m) $ orders being generated in the high harmonics. Figure \ref{high_order_mode} shows the electric field distribution in the $ x-y $ plane for the third and fifth harmonics driven by radially polarized lasers. The black dashed lines denote the cut positions along the $ x $-plane in Fig. \ref{hhg_nor}. The field pattern clearly demonstrates the intensity modulation is due to higher $ (l,m) $ transverse modes but not wavefront distortion. This is in contrast to the driving laser pulses and the case of relativistic vortex HHG by Laguerre-Gaussian driving lasers\cite{zhang_generation_2015}, but is similar to the results observed in vortex HHG from gases\cite{geneaux_radial_2017,jin_phase-matching_2020}. In the later case, the appearance of higher-order radial modes is attributed to spatial phase acquired by the HHG process, which might have some implications for the origin of the extra radial modes in our case.

Figure \ref{trans_profile}(a) shows the radially polarized laser driven harmonic peak intensity as a function of harmonic order $ n $. The spectral scaling is close to the $ n^{-8/3} $ law given by the $ \gamma $-spike model of HHG based on ROM\cite{baeva_theory_2006}, suggesting that a relatively high conversion efficiency may be achieved. Although the scheme of normal incidence is avoided in real experiments to prevent light reflection from damaging the optical components, small-angle incidence should work and give HHG results close to that of normal incidence.

\subsection{Vector attosecond pulses}
\begin{figure*}[htbp]
	\centering
	\includegraphics[width=0.8\textwidth
	]{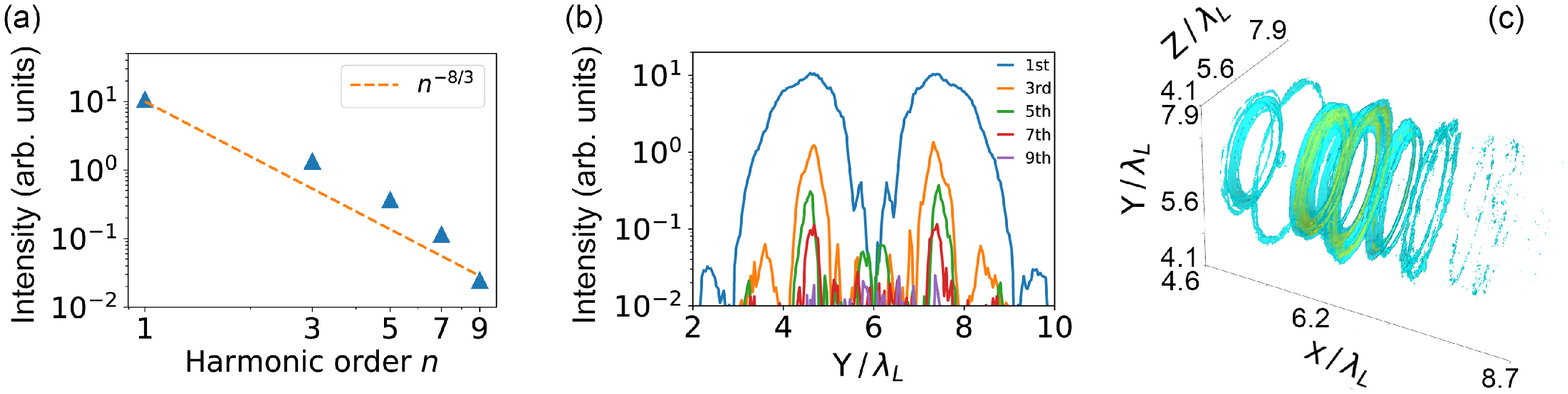}
	\caption{\label{trans_profile} (a) The peak intensity of high harmonics as a function of harmonic order $ n $. The dashed orange line shows the $ n^{-8/3} $ spectral scaling law given by the $ \gamma $-spike model of HHG based on relativistically oscillating mirrors. (b) 1D (lineout along the $ z=0 $ axis) transverse intensity profiles of the fundamental field and the third, fifth, seventh, and ninth harmonics. (c) Isosurface distribution of the high harmonics after spectral filtering by selecting the 3rd-23rd orders, showing an attosecond pulse train structure. These results are corresponding to the case of radially polarized driving lasers.}
\end{figure*}

Another advantage of generating XUV/X-ray vector beams through HHG is that high harmonics are naturally emitted in the form of attosecond pulse in the time domain. The 1D lineout (cut along the $ z=0 $ axis) intensity profiles for different harmonic orders driven by radially polarized lasers are shown in Fig. \ref{trans_profile}(b). It is seen that the transverse radius of the mainlobe decreases with increasing the harmonic order, while the peak position are approximately the same. This spatial overlap of different harmonic orders allows the synthesis of attoseond pulses. Figure \ref{trans_profile}(c) shows the isosurface profiles of the high harmonics driven by radially polarized lasers, which are obtained after spectral filtering by selecting the 3rd-23rd harmonic orders. Field structure of an attosecond pulse train can be clearly seen. Combined with the already demonstrated characteristic of radial polarization, these intense attosecond high harmonic vector beams may open up many new applications such as probing ultrafast magnetic or spin dynamics.

\subsection{Vector beam with OAM}
\begin{figure*}[htbp]
	\centering
	\includegraphics[width=0.85\textwidth
	]{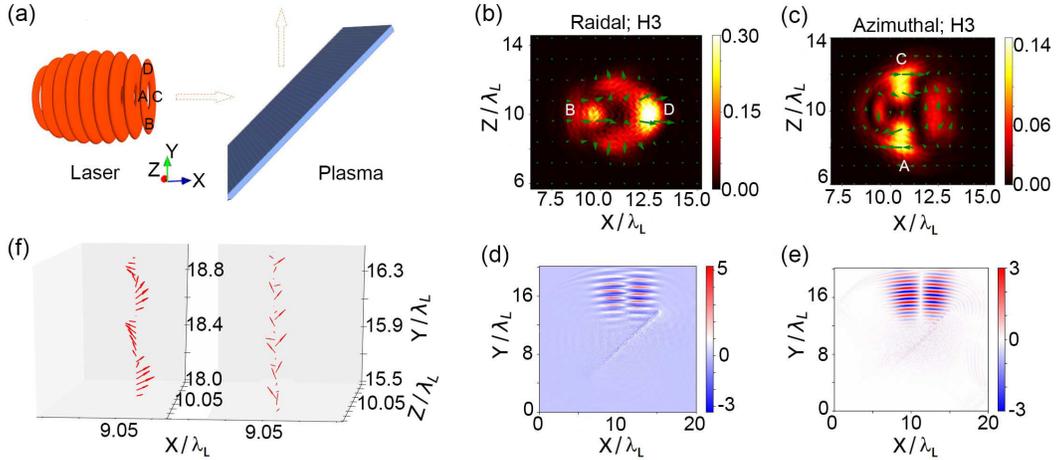}
	\caption{\label{ob_scheme} (a) Sketch of simulation setup for oblique incidence. The incident vector laser beams propagate along the $ x $-axis. The plasma slab is rotated clockwise by $ 45^{\circ} $ around the $ z $-axis. Then the reflected pulses propagate along the $ y $-axis. (b-c) The intensity pattern (color scale) and spatial polarization distribution (green arrows) of the third harmonic driven by the (b) radially and (c) azimuthally polarized laser beams observed in the transverse $ x-z $ plane at $ y=16.2\lambda_L $. (d-e) The electric field distribution of the reflected pulses driven by the (d) radially and (e) azimuthally polarized laser beams observed in the $ x-y $ plane at $ z=0 $. (f) The local Poynting vectors of the reflected beams (left) and third harmonics (right) driven by radially polarized lasers, showing vector beams carrying a nonzero orbital angular momentum having been generated.}
\end{figure*}

Next we investigate the case of oblique incidence. For convenience of computation and analysis, we set the propagation axis of the incident lasers to coincide with the $ x $-axis, but the incident surface of plasma slab (with a uniform density of $ 40n_c $) to be rotated $ 45^{\circ} $ clockwise around the $ z $-axis, so that the reflected pulses propagate along the $ y $-axis (see Fig. \ref{ob_scheme}(a)). The harmonic intensity pattern and spatial polarization distribution in the transverse $ x-z $ plane at $ y=16.2\lambda_L $ for the 3rd (H3) harmonic driven by radially and azimuthally polarized lasers are shown in Fig. \ref{ob_scheme}(b) and Fig. \ref{ob_scheme}(b), respectively.  For each case, two bright harmonic spots separated in the transverse plane can be observed from the intensity pattern, as a result of partial $ p $- and $ s $-polarization in the electric field structure of the laser pulses with respect to the plasma surfaces, in consistent with the results reported by Za\"{i}m et al\cite{zaim_interaction_2020}. Specifically, for the case of radially polarized driving lasers, the B and D parts (see Fig. \ref{ob_scheme}(a)) of the lasers are $ p $-polarized (i.e., the electric field vectors are perpendicular to the plasma surface), while the A and C parts of the lasers are $ s $-polarized (i.e., the electric field vectors are parallel to the plasma surface). $ p $-polarization is more efficient than $ s $-polarization in driving oscillation of the relativistic plasma mirrors, thus the resultant HHG is more intense in the regions corresponding to B and D (see Fig. \ref{ob_scheme}(b)). Harmonic intensity in region D is higher than that in region B, which may be explained by lasers in part D arriving at the plasma surface at a later time when the plasmas are already heated for a while. In comparison, for the case of azimuthally polarized driving lasers, the A and C parts of the lasers are $ p $-polarized, and thus the harmonics in the corresponding regions are brighter. The harmonic intensity in the A and C regions are about the same level, since both parts of the lasers illuminate the plasma surfaces about the same time. 

For the spatial polarization distribution, we see from the green arrows that the vector light feature is also transferred to the harmonic beams. This shows the potential advantages to generate even brighter vector high harmonic beams, as oblique incidence is more efficient in HHG than normal incidence. One may note that in the middle left and middle right regions in Figs. \ref{ob_scheme}(b)-(c), the phases of electric fields are somewhat opposite to form a regular radial or azimuthal polarization together with the other parts of electric field in the shown plane. This is a result of wavefront distortion caused by oblique incidence. In the $ x-y $ plane, the lower and upper parts of the vector laser beam arrive at the plasma surface at different times. The time delay in field reflection leads to a phase shift in the electric field. Figures \ref{ob_scheme}(d)-(e) show the electric field distribution of the reflected pulses in the $ x-y $ plane at $ z=0 $ for the cases of radially and azimuthally polarized driving lasers, respectively. We see that the reflected electric fields indeed exhibit phase shift between the two sections of the hollow laser profiles along each $ y=constant $ plane. 

Due to the phase shift, the high harmonics under oblique incidence are still vector beams, but no longer exactly the same type of vector beam as that of the fundamental laser pulse. Instead, they present interesting polarization distribution in the transverse plane, which may be tuned by the laser incidence angle and thus the phase shift. This offers a way to generate novel kinds of vector beams. 

More interestingly, the different field phases imply that vector harmonic beams carrying OAM may be generated. As pointed out by Kotlyar \textit{et al}., the linear combination of HG beams with a phase delay can have a nonzero OAM\cite{Kotlyar_Hermite-Gaussian_2014}. Such a beam is lack of radial symmetry and preserves its structure during propagation. Figure \ref{ob_scheme}(f) shows the local Poynting vector ($ \textbf{S}=\textbf{E}\times \textbf{B} $) of the reflected pulse and third harmonics for the case of radially polarized driving lasers. The spiral structure of the Poynting vector about the optical axis in space evidently demonstrates the beams indeed have nonzero OAM. Different from previous studies of vortex HHG with linear\cite{zhang_generation_2015} or circular polarization\cite{wang_intense_2019}, here the high harmonic beam is simultaneously phase (optical vortices) and polarization structured (vector beams). The vortices and vector structures may both be tuned by changing the incidence angle.

\subsection{High resolution 2D simulations}

\begin{figure}[htbp]
	\centering
	\includegraphics[width=0.5\textwidth
	]{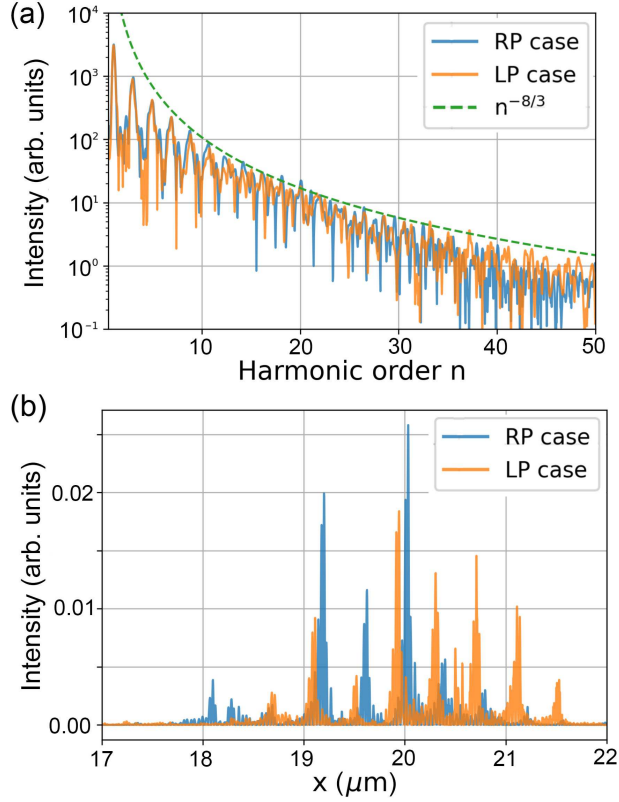}
	\caption{\label{cyl_1d} 2D cylindrical PIC simulation results. (a) Spectra of the reflected pulses for the radially polarized (RP) (2D cylindrical simulation) and linearly polarized (LP) (1D simulation) driving laser cases. (b) Intensity profiles of the reflected pulses after spectral filtering by selecting the harmonics with wavelength in the range of 8-80 nm (10th-100th harmonic orders). The profiles of the RP case are taken along $r=w_0/\sqrt{2}$.}
\end{figure}

Finally, we address the generation of vector harmonic beams in higher frequency range. Although 3D PIC simulations retain all the degrees of freedom of the physical system investigated here, the temporal resolution and spectral range of the simulation are limited by the computational resources. For the normal incidence case, the physical system has a perfect axial-symmetry. Thus 2D cylindrical PIC simulations are very suitable and efficient for such a system\cite{lifschitz_particle_2009}. To investigate the spectral and temporal distribution of the reflected HHG with much higher resolutions, we perform additional 2D cylindrical PIC simulations for the radially polarized (RP) driving laser case. The simulation box has 15000$\times$500 cells, corresponding to a cylindrical volume with $37.5 \lambda_L$ in the laser propagation direction $x$ and $62.5 \lambda_L$ in the radial direction $r$. The laser has a normalized amplitude of $a_0=5$. The waist of the laser is $w_0=12.5\lambda_L$ and the pulse duration is $\tau_0=2.65T_0$. The plasma slab has a thickness of 0.125$\lambda_L$ and density of 11.5$n_c$. The preplasma has a scale length of $0.25\lambda_L$. Spectrum of the reflected RP pulse is shown in Fig. \ref{cyl_1d}(a). High harmonics extending to about 40th harmonic order can be clearly observed. For comparison, harmonic spectrum driven by a linearly polarized (LP) laser with $a_0=2.14$ obtained from 1D PIC simulations is also shown. Both spectral scalings are close to the $n^{-8/3}$ law given by the ROM model. These results demonstrate the potential of generating vector beams with high brightness extending to the XUV and X-ray spectral region. Figure \ref{cyl_1d}(b) shows the intensity profiles of the reflected pulses after spectral filtering by selecting the 10th-100th harmonic orders (8-80 nm). A train of attosecond pulses with duration of a few hundred attoseconds is generated by the ROM mechanism. The intensity of the attosecond pulses for the RP and LP cases is slightly different, which may have contributions from multidimensional effects like focusing and defocusing from curved reflecting planes.


\section{Conclusions}
In conclusion, we numerically demonstrate the generation of coherent vector beams in the XUV regime by means of HHG from relativistic plasma mirrors. The vector harmonic beams have the advantages of high brightness and ultrashort duration down to the attosecond scale. The frequency range can be extended to higher photon energies up to the soft x-ray regime. Vector high harmonic beams with nonzero OAM can possibly be generated under oblique incidence. Such novel light sources offer new opportunities in a number of applications ranging from imaging with high spatial and temporal resolution, and magnetic spectroscopy, to particle manipulation.


\section*{acknowledgments}
This work was supported in part by the National Key Laboratory of Shock Wave and Detonation Physics (JCKYS2020212015,JCKYS2019212003), National Natural Science Foundation of China (11705185), and the Fundamental Research Funds for the Central Universities (YJ202025).

\bibliography{ref_vechhg}
\end{document}